\documentclass[10.75pt,preprint]{article}
\usepackage[english]{babel}
\usepackage{amsmath,amssymb,graphicx,calc,capt-of,ifthen,theorem,pifont,amsfonts}

\begin{document}

\title{$\tilde{\delta}$ Gravity and Schwarzschild Solution.}
\author{Jorge Alfaro\footnote{Facultad de F\'isica, Pontificia Universidad Cat\'olica de Chile. Casilla 306, Santiago, Chile. jalfaro@uc.cl.} and Pablo Gonz\'alez\footnote{Departamento de F\'isica, FCFM, Universidad de Chile. Blanco Encalada 2008, Santiago, Chile. pgonzalez@ing.uchile.cl (Current Affiliation).} \footnote{Facultad de F\'isica, Pontificia Universidad Cat\'olica de Chile. Casilla 306, Santiago, Chile. pegonza2@uc.cl.}}
\maketitle

\begin{abstract}
$\tilde{\delta}$ Gravity is a gravitational field model, where the geometry is governed by two symmetric tensors, $g_{\mu \nu}$ and $\tilde{g}_{\mu \nu}$, and new matter fields ($\tilde{\delta}$ Matter fields) are added to the original matter fields. These new components appear motivated by a new symmetry, called $\tilde{\delta}$ symmetry. In previous works, the model is used to explain the expansion of the Universe without Dark Energy and use $\tilde{\delta}$ Matter as a source of Dark Matter. In this paper, we will developed an initial study of Schwarzschild geometry in $\tilde{\delta}$ Gravity to complement the Dark Matter analysis and introduce other phenomena. We will get a modified deflection of light produced by the sun, the perihelion precession and black hole solution.
\end{abstract}

\section*{Introduction.}

Recent discoveries in cosmology have revealed that most part of matter is in the form of unknown matter, Dark Matter \cite{DM DE 1}-\cite{DM New 8}, and that the dynamics of the expansion of the Universe is governed by a mysterious component that accelerates its expansion, the so called Dark Energy \cite{DM DE 2}-\cite{DM DE 4}. That is the Dark Sector. Although General Relativity (GR) is able to accommodate the Dark Sector, its interpretation in terms of fundamental theories of elementary particles is problematic \cite{DM DE 5}.\\

In relation to Dark Energy, many experiments are been carried out to determine its nature \cite{DM DE 1}, however its detection has been problematic. In a galactic scale, Dark Matter produces an anomalous rotation velocity where it is relatively constant far from the center of the galaxy \cite{DM New 1}-\cite{DM New 8}, and a lot of alternative models, where a modification to gravity is introduced, have been developed to explain this effect. For instance, a explanation based on the modification of the dynamics for small accelerations cannot be ruled out \cite{DM DE 6,DM DE 7}. On the other side, the accelerated expansion of the universe can be explained if a small cosmological constant ($\Lambda$) is present. However $\Lambda$ is too small to be generated in quantum field theory (QFT) models, because $\Lambda$ is the vacuum energy, which is usually predicted to be very large \cite{lambda problem}.\\

In GR, the Schwarzschild metric is used to study many phenomena. For instance, we have the deflection of light by gravitational lensing, obtained with an excellent precision in GR \cite{Deflect}. In the same way, we have the perihelion precession used to explain the anomalous orbital trajectory of mercury. All these phenomena were really important to verify GR, so any modification of gravity must be very close to GR in the solar system scale, where these effects have been studied. On the other side, it can be used to study a Schwarzschild black hole. In a modified gravity model, we could obtain new effects in the structure of black holes.\\

In \cite{delta gravity}, we presented a model of gravitation, called $\tilde{\delta}$ Gravity, that is very similar to GR, but works different in a quantum level. The foundation of $\tilde{\delta}$ Gravity considers two different points. The first is that GR is finite on shell at one loop in vacuum {\cite{tHooft}}, so renormalization is not necessary at this level. The second is the $\tilde{\delta}$ gauge theories (DGT) originally presented in {\cite{Alfaro 2,Alfaro 3}}, where the main properties are: (a) A new kind of field $\tilde{\phi}_{I}$ is introduced, different from the original set $\phi_{I}$. (b) The classical equations of motion of $\phi_{I}$ are satisfied even in the full quantum theory. (c) The model lives at one loop. (d) The action is obtained through the extension of the original gauge symmetry of the model, introducing an extra symmetry that we call $\tilde{\delta}$ symmetry, since it is formally obtained as the variation of the original symmetry. When we apply this prescription to GR, we obtain $\tilde{\delta}$ Gravity. For these reasons, the original motivation was to develop the quantum properties of this model (See \cite{delta gravity}). But now, we prefer to emphasize the use of $\tilde{\delta}$ Gravity as an effective model of gravitation and explore its phenomenological predictions.\\

In other works, we presented a truncated version of $\tilde{\delta}$ Gravity to explain the accelerated expansion of the universe without Dark Energy \cite{DG DE,Paper DE}. The $\tilde{\delta}$ symmetry was fixed in different ways in order to simplify the analysis of the model, however $\tilde{\delta}$ Matter was ignored in the process. After, in \cite{Paper 1}, we presented a $\tilde{\delta}$ Gravity version, where $\tilde{\delta}$ symmetry and $\tilde{\delta}$ Matter are preserved. In that case, the accelerated expansion can be explained in the same way and additionally we have a new component of matter as Dark Matter candidate. Besides, we guaranteed that the special properties of $\tilde{\delta}$ Theories previously mentioned are preserved.\\

Later, in \cite{Paper 2}, the Non-Relativistic limit was studied to explain Dark Matter with $\tilde{\delta}$ Matter in different scales, using some realistic density profiles as Einasto and Navarro-Frenk-White (NFW) profiles. We saw that $\tilde{\delta}$ Matter effect is not related to the scale, but rather to the behavior of the distribution of ordinary matter. Additionally, $\tilde{\delta}$ Matter produce an amplifying effect in the total mass and rotation velocity in a galaxy. So, the Dark Matter effect could be explained with a considerably less quantity of ordinary Dark Matter, explaining its extremely problematic detection. Besides, in \cite{Paper 1}, we obtained that the non-relativistic $\tilde{\delta}$ Matter quantity in the present at cosmological level is 23\% of the ordinary non-relativistic matter, where Dark Matter is included, implying that Dark Matter is in part $\tilde{\delta}$ Matter.\\

In this paper, we will see that $\tilde{\delta}$ Gravity agrees with GR at the classical level far from the sources. In particular, the causal structure of $\tilde{\delta}$ Gravity in vacuum is the same as in GR. Besides, the $\tilde{\delta}$ Matter is negligible to Solar Mass, but important effects appear when really massive object, as Black Holes, are taken into account. In \textbf{Section \ref{Sec: delta Gravity}}, we will present the $\tilde{\delta}$ Gravity action that is invariant under extended general coordinate transformation. We will find the equations of motion of this action. We will see that the Einstein's equations continue to be valid and obtain a new equation for $\tilde{g}_{\mu \nu}$. In these equations, two energy momentum tensors, $T_{\mu \nu}$ and $\tilde{T}_{\mu \nu}$, are defined. Additionally, we will derive the equation of motion for a test particle. We will distinguish the massive case, where the equation is not a geodesic, and the massless case, where a null geodesic with an effective metric governs the evolution of the trajectory. In \textbf{Section \ref{Sec: Schwarzschild Case}}, we will study the Schwarzschild Case. The equations of motion of $g_{\mu \nu}$ and $\tilde{g}_{\mu \nu}$ will be solved in vacuum with appropriate boundary conditions. Then, we will use this solution to compute the deflection of light and analyze the perihelion precession. We have to guarantee that these results are very close to GR, unless we consider highly massive object. Finally, we will look for a connection with Dark Matter and we will introduce a black hole analysis.\\

Before continuing, we want to introduce a word of caution. In what follows, we want to study $\tilde{\delta}$ Gravity as a \textit{classical effective model}. This means to approach the problem from the phenomenological side instead of neglecting it \textit{a priori} because it does not satisfy yet all the properties of a fundamental quantum theory. In a cosmological level, the observations indicate that a phantom component is compatible with most classical tests of cosmology based on current data \cite{phantom 1}-\cite{phantom 6}. The nature of the Dark Sector is such an important and difficult cosmological problem that cosmologists do not expect to find a fundamental solution of it in one stroke and are open to explore new possibilities. Now, the \textit{phantom problem} is being studied in this moment and the results will be presented in a future work.\\

Additionally, it should be remarked that $\tilde{\delta}$ Gravity is not a metric model of gravity because massive particles do not move on geodesics. Only massless particles move on null geodesics of a linear combination of both tensor fields. Additionally, it is important to notice that we will work with the $\tilde{\delta}$ modification for GR, based on the Einstein-Hilbert theory. From now on, we will refer to this model as $\tilde{\delta}$ Gravity.\\


\section{\label{Sec: delta Gravity}$\tilde{\delta}$ Gravity.}

Using the prescription given in \textbf{Appendix A}, we will present the action of $\tilde{\delta}$ Gravity and then we will derive the equations of motion. Additionally, we will study the test particle action separately for massive and massless particles. We will see the geometry in $\tilde{\delta}$ Gravity produces different effects in both cases.\\

\subsection{\label{SubSec: Equations of Motion}Equations of Motion:}

To obtain the action of $\tilde{\delta}$ Gravity, we will consider the Einstein-Hilbert Action:

\begin{eqnarray}
\label{EH action}
S_0 = \int d^4x \sqrt{-g} \left(\frac{R}{2\kappa} + L_M\right),
\end{eqnarray}

where $L_M = L_M(\phi_I,\partial_{\mu}\phi_I)$ is the lagrangian of the matter fields $\phi_I$. Using (\ref{Action}) from \textbf{Appendix A}, this action becomes:

\begin{eqnarray}
\label{grav action}
S = \int d^4x \sqrt{-g} \left(\frac{R}{2\kappa} + L_M - \frac{1}{2\kappa}\left(G^{\alpha \beta} - \kappa T^{\alpha \beta}\right)\tilde{g}_{\alpha \beta} + \tilde{L}_M\right),
\end{eqnarray}

where $\kappa = \frac{8 \pi G}{c^2}$, $\tilde{g}_{\mu \nu} = \tilde{\delta}g_{\mu \nu}$ and:

\begin{eqnarray}
\label{EM Tensor}
T^{\mu \nu} = \frac{2}{\sqrt{-g}} \frac{\delta}{\delta g_{\mu \nu}}\left[\sqrt{-g} L_M\right] \\
\label{tilde L matter}
\tilde{L}_M = \tilde{\phi}_I\frac{\delta L_M}{\delta \phi_I} + (\partial_{\mu}\tilde{\phi}_I)\frac{\delta L_M}{\delta (\partial_{\mu}\phi_I)},
\end{eqnarray}

where $\tilde{\phi}_I = \tilde{\delta}\phi_I$ are the $\tilde{\delta}$ Matter fields. These matter components are the fundamental difference with \cite{Paper DE}. From this action, we can see that the Einstein's equations are preserved and an additional equation for $\tilde{g}_{\mu \nu}$ is obtained. So, the equations of motion are:

\begin{eqnarray}
\label{Einst Eq} G^{\mu \nu} &=& \kappa T^{\mu \nu} \\
\label{tilde Eq} F^{(\mu \nu) (\alpha \beta) \rho
\lambda} D_{\rho} D_{\lambda} \tilde{g}_{\alpha \beta} + \frac{1}{2}g^{\mu \nu}R^{\alpha \beta}\tilde{g}_{\alpha \beta} - \frac{1}{2}\tilde{g}^{\mu \nu}R &=& \kappa\tilde{T}^{\mu \nu}.
\end{eqnarray}

with:

\begin{eqnarray}
\label{F}
F^{(\mu \nu) (\alpha \beta) \rho \lambda} &=& P^{((\rho
\mu) (\alpha \beta))}g^{\nu \lambda} + P^{((\rho \nu) (\alpha
\beta))}g^{\mu \lambda} - P^{((\mu \nu) (\alpha \beta))}g^{\rho
\lambda} - P^{((\rho \lambda) (\alpha \beta))}g^{\mu \nu} \nonumber \\
P^{((\alpha \beta)(\mu \nu))} &=& \frac{1}{4}\left(g^{\alpha
\mu}g^{\beta \nu} + g^{\alpha \nu}g^{\beta \mu} - g^{\alpha
\beta}g^{\mu \nu}\right),
\end{eqnarray}

where $(\mu \nu)$ denotes that $\mu$ and $\nu$ are in a totally symmetric combination. An important fact to notice is that our equations are of second order in derivatives which is needed to preserve causality. Finally, from \textbf{Appendix A}, we have that the action (\ref{grav action}) is invariant under (\ref{trans g}) and (\ref{trans gt}). This means that two conservation rules are satisfied. They are:

\begin{eqnarray}
\label{Conserv T}
D_{\nu}T^{\mu \nu} &=& 0 \\
\label{Conserv tilde T}
D_{\nu}\tilde{T}^{\mu \nu} &=& \frac{1}{2}T^{\alpha \beta}D^{\mu}\tilde{g}_{\alpha \beta} - \frac{1}{2}T^{\mu \beta} D_{\beta}\tilde{g}^{\alpha}_{\alpha} + D_{\beta}(\tilde{g}^{\beta}_{\alpha}T^{\alpha \mu}).
\end{eqnarray}

In conclusion, we have that the original equation in the Einstein-Hilbert Action, given by (\ref{Einst Eq}) and (\ref{Conserv T}), are preserved. This is one of the principal properties of $\tilde{\delta}$ Theories. Additionally, we have two new equations, (\ref{tilde Eq}) and (\ref{Conserv tilde T}), to obtain the solution of $\tilde{g}_{\mu \nu}$ and $\tilde{\delta}$ Matter respectively. These components will produce new contributions to some phenomena, usually taken into account in (\ref{Einst Eq}) and (\ref{Conserv T}) in GR. For example, in \cite{Paper 1}, a universe without a cosmological constant is considered, so the accelerated expansion of the universe is produced by the additional effect of $\tilde{\delta}$ components. In this paper, we will introduce the Schwarzschild solution to study any modification produced by if $\tilde{\delta}$ Gravity.\\

Now, to complete our theoretical framework, we will show how our two gravitational components, $g_{\mu \nu}$ and $\tilde{g}_{\mu \nu}$, interact with a test particle.\\

\subsection{\label{SubSec: Test Particle}Test Particle:}

To understand how the new fields affect the trajectory of a particle, we need to study the Test Particle Action. However, it must be analyzed separately for massive and massless particles. The first discussion of this issue in $\tilde{\delta}$ Gravity is in \cite{DG DE}.\\

\subsubsection{\label{SubSubSec: Massive Particles}Massive Particles:}

In GR, the action for a test particle is given by:

\begin{eqnarray}
\label{Geo Action 0}
S_0[\dot{x},g] = - m \int dt \sqrt{-g_{\mu \nu}\dot{x}^{\mu}\dot{x}^{\nu}},
\end{eqnarray}

with $\dot{x}^{\mu} = \frac{d x^{\mu}}{d t}$. This action is invariant under reparametrizations, $t' = t - \epsilon(t)$. In the infinitesimal form is:

\begin{eqnarray}
\label{reparametr}
\delta_R x^{\mu} &=& \dot{x}^{\mu}\epsilon.
\end{eqnarray}

In $\tilde{\delta}$ Gravity, the action is always modified using (\ref{Action}) from \textbf{Appendix A}. So, applying it to (\ref{Geo Action 0}), the new test particle action is:

\begin{eqnarray}
\label{Geo Action 01}
S[\dot{x},y,g,\tilde{g}] &=& - m \int dt \sqrt{-g_{\mu \nu}\dot{x}^{\mu}\dot{x}^{\nu}} + \frac{m}{2} \int dt \left(\frac{\tilde{g}_{\mu \nu}\dot{x}^{\mu}\dot{x}^{\nu}+g_{\mu \nu,\rho}y^{\rho}\dot{x}^{\mu}\dot{x}^{\nu}+2g_{\mu \nu}\dot{x}^{\mu}\dot{y}^{\nu}}{\sqrt{-g_{\mu \nu}\dot{x}^{\mu}\dot{x}^{\nu}}}\right) \nonumber \\
&=& m \int dt \left(\frac{\left(g_{\mu \nu} + \frac{1}{2}\tilde{g}_{\mu \nu}\right)\dot{x}^{\mu}\dot{x}^{\nu} + \frac{1}{2}(2g_{\mu \nu}\dot{y}^{\mu}\dot{x}^{\nu} + g_{\mu \nu, \rho}y^{\rho}\dot{x}^{\mu}\dot{x}^{\nu})}{\sqrt{-g_{\alpha \beta}\dot{x}^{\alpha}\dot{x}^{\beta}}}\right),
\end{eqnarray}

where we have defined $y^{\mu} = \tilde{\delta}x^{\mu}$ and we used that $g_{\mu \nu} = g_{\mu \nu}(x)$, so $\tilde{\delta}g_{\mu \nu} = \tilde{g}_{\mu \nu} + g_{\mu \nu, \rho}y^{\rho}$. Naturally, this action is invariant under reparametrization transformations, given by (\ref{reparametr}), plus $\tilde{\delta}$ reparametrization transformations:

\begin{eqnarray}
\label{reparametr plus}
\delta_R y^{\mu} &=& \dot{y}^{\mu}\epsilon + \dot{x}^{\mu}\tilde{\epsilon},
\end{eqnarray}

just like it is shown in (\ref{tilde trans general}). The presence of $y^{\mu}$ suggests additional coordinates, but our model just live in four dimensions, given by $x^{\mu}$. Actually, $y^{\mu}$ can be gauged away using the extra symmetry corresponding to $\tilde{\epsilon}$ in equation (\ref{reparametr plus}), imposing the gauge condition $2g_{\mu \nu}\dot{y}^{\mu}\dot{x}^{\nu} + g_{\mu \nu, \rho}y^{\rho}\dot{x}^{\mu}\dot{x}^{\nu} = 0$. However, the extended general coordinate transformations as well as the usual reparametrizations, given by (\ref{reparametr}), are still preserved. Then, (\ref{Geo Action 01}) can be reduced to:

\begin{eqnarray}
\label{Geo Action}
S[\dot{x},g,\tilde{g}] = m \int dt \left(\frac{\left(g_{\mu \nu} + \frac{1}{2}\tilde{g}_{\mu \nu}\right)\dot{x}^{\mu}\dot{x}^{\nu}}{\sqrt{-g_{\alpha \beta}\dot{x}^{\alpha}\dot{x}^{\beta}}}\right).
\end{eqnarray}

Notice that the test particle action in Minkowski space is recovered if we used the boundary conditions, given by $g_{\mu \nu} \sim \eta_{\mu \nu}$ and $\tilde{g}_{\mu \nu} \sim 0$, to be equal to GR. So, this action for a test particle in a gravitational field will be considered as the starting point for the physical interpretation of the geometry in $\tilde{\delta}$ Gravity. Now, the trajectory of massive test particles is given by the equation of motion of $x^{\mu}$. This equation say us that $g_{\mu \nu}\dot{x}^{\mu}\dot{x}^{\nu} = cte$, just like GR. Now, if we choose $t$ equal to the proper time, then $g_{\mu \nu}\dot{x}^{\mu}\dot{x}^{\nu} = -1$  and the equation of motion is reduced in this case to:

\begin{eqnarray}
\label{geodesics m}
\hat{g}_{\mu \nu} \ddot{x}^{\nu} + \hat{\Gamma}_{\mu \alpha \beta} \dot{x}^{\alpha} \dot{x}^{\beta} = \frac{1}{4}\tilde{K}_{,\mu},
\end{eqnarray}

with:

\begin{eqnarray}
\hat{\Gamma}_{\mu \alpha \beta} &=& \frac{1}{2}(\hat{g}_{\mu \alpha , \beta} + \hat{g}_{\beta \mu , \alpha} - \hat{g}_{\alpha \beta ,
\mu})\nonumber \\
\hat{g}_{\alpha \beta} &=& \left(1+\frac{1}{2} \tilde{K}\right)g_{\alpha \beta} + \tilde{g}_{\alpha \beta}
\nonumber \\
\tilde{K} &=& \tilde{g}_{\alpha \beta} \dot{x}^{\alpha} \dot{x}^{\beta}. \nonumber
\end{eqnarray}

This equation of motion is independent of the mass of the particle, so all particles will fall with the same acceleration. On the other side, the equation (\ref{geodesics m}) is a second order equation, but it is not a classical geodesic, because we have additional terms and an effective metric can not be defined.\\

\subsubsection{\label{SubSubSec: Massless Particles}Massless Particles:}

We saw that the free trajectory for a massive particle is an anomalous geodesic given by (\ref{geodesics m}). However, this equation is useless for massless particles, because (\ref{Geo Action 0}) is null when $m=0$. To solve this problem, it is a common practice to start from the action \cite{massless geo}:

\begin{eqnarray}
\label{Geo Action 0 Lagr}
S_0[\dot{x},g,v] = \frac{1}{2} \int dt \left(vm^2 - v^{-1}g_{\mu \nu}\dot{x}^{\mu}\dot{x}^{\nu}\right),
\end{eqnarray}

where $v$ is an auxiliary field, which transforms under reparametrizations as:

\begin{eqnarray}
\label{v}
v'(t')&=&\frac{d t}{d t'}v(t).
\end{eqnarray}

From (\ref{Geo Action 0 Lagr}), we can obtain the equation of motion for $v$:

\begin{eqnarray}
\label{v eq}
v = - \frac{\sqrt{-g_{\mu \nu}\dot{x}^{\mu}\dot{x}^{\nu}}}{m}.
\end{eqnarray}

We see from (\ref{v}) that the gauge $v=constant$ can be fixed, so in GR the proper time $\sqrt{-g_{\mu \nu}\dot{x}^{\mu}\dot{x}^{\nu}}$ remains constant along the path. Now, if we substitute (\ref{v eq}) in (\ref{Geo Action 0 Lagr}), we recover (\ref{Geo Action 0}). This means (\ref{Geo Action 0 Lagr}) is equivalent to (\ref{Geo Action 0}), but additionally includes the massless case.\\

In our case, a suitable action, similar to (\ref{Geo Action 0 Lagr}), is:

\begin{eqnarray}
\label{Geo Action Lagr 2}
S[\dot{x},g,\tilde{g},v] = \int dt \left(m^2v - \frac{\left(g_{\mu \nu} + \tilde{g}_{\mu \nu}\right)\dot{x}^{\mu}\dot{x}^{\nu}}{4v} + \frac{m^2 v^3}{4g_{\alpha \beta}\dot{x}^{\alpha}\dot{x}^{\beta}}\left(m^2 + v^{-2} \tilde{g}_{\mu \nu}\dot{x}^{\mu}\dot{x}^{\nu}\right)\right).
\end{eqnarray}

In fact, in $\tilde{\delta}$ Gravity the equation of $v$ is still (\ref{v eq}). Thus, we can fix the gauge $v=constant$, to remain constant the quantity $\sqrt{-g_{\mu \nu}\dot{x}^{\mu}\dot{x}^{\nu}}$ along the path, just like GR. This is the reason why we can choose $t$ as the proper time in $\tilde{\delta}$ Gravity too. Additionally, if we replace \label{v} in (\ref{Geo Action Lagr 2}), we obtain the massive test particle action given by (\ref{Geo Action}). But now, we can study the massless case.\\

If we evaluate $m = 0$ in (\ref{Geo Action 0 Lagr}) and (\ref{Geo Action Lagr 2}), we can compare GR and $\tilde{\delta}$ Gravity respectively. They are:

\begin{eqnarray}
\label{Geo Action 0 foton}
S^{(m=0)}_0[\dot{x},g,v] &=& - \frac{1}{2} \int dt v^{-1}g_{\mu \nu}\dot{x}^{\mu}\dot{x}^{\nu} \\
\label{Geo Action foton}
S^{(m=0)}[\dot{x},g,\tilde{g},v] &=& - \frac{1}{4} \int dt v^{-1}\mathbf{g}_{\mu \nu}\dot{x}^{\mu}\dot{x}^{\nu},
\end{eqnarray}

with $\mathbf{g}_{\mu \nu} = g_{\mu \nu} + \tilde{g}_{\mu \nu}$. In both cases, the equation of motion for $v$ implies that a massless particle move in a null-geodesic. In the usual case we have $g_{\mu \nu}\dot{x}^{\mu}\dot{x}^{\nu} = 0$. However, in our model the null-geodesic is given by $\mathbf{g}_{\mu \nu}\dot{x}^{\mu}\dot{x}^{\nu} = 0$, so the trajectory obey a geometry defined by an specifical combination of $g_{\mu \nu}$ and $\tilde{g}_{\mu \nu}$, $\mathbf{g}_{\mu \nu} = g_{\mu \nu} + \tilde{g}_{\mu \nu}$. The equation of motion for the path of a test massless particle is given by:

\begin{eqnarray}
\label{geodesics null}
\mathbf{g}_{\mu \nu} \ddot{x}^{\nu} + \mathbf{\Gamma}_{\mu \alpha \beta} \dot{x}^{\alpha} \dot{x}^{\beta} = 0 \\
\mathbf{g}_{\mu \nu}\dot{x}^{\mu}\dot{x}^{\nu} = 0, \nonumber
\end{eqnarray}

with:

\begin{eqnarray}
\mathbf{\Gamma}_{\mu \alpha \beta} = \frac{1}{2}(\mathbf{g}_{\mu \alpha , \beta} + \mathbf{g}_{\beta \mu , \alpha} - \mathbf{g}_{\alpha \beta, \mu}). \nonumber
\end{eqnarray}

It is important to observe that the proper time must be defined in terms of massive particles. The equation of motion for massive particles satisfies the important property of preserving the form of the proper time in a particle in free fall. Notice that in our case the quantity that is constant using the equation of motion for massive particles, derived from (\ref{geodesics m}), is $g_{\mu \nu}\dot{x}^{\mu}\dot{x}^{\nu} = -1$. This single out this definition of proper time and not other. So, we must define proper time using the original metric $g_{\mu \nu}$. That is:

\begin{eqnarray}
\label{proper time}
g_{\mu \nu}\left(\frac{1}{c}\frac{dx^{\mu}}{d\tau}\right)\left(\frac{1}{c}\frac{dx^{\nu}}{d\tau}\right) = - 1 \nonumber \\
\Rightarrow d \tau = \frac{1}{c} \sqrt{-g_{\mu \nu}dx^{\mu}dx^{\nu}} \rightarrow \sqrt{-g_{0 0}} dt.
\end{eqnarray}

From here, we can see that $g_{00}<0$. On the other side, if we consider the motion of light rays along infinitesimally near trajectories, from (\ref{geodesics null}) and (\ref{proper time}) we get the three-dimensional metric (See \cite{DG DE,Landau}):

\begin{eqnarray}
\label{tri metric}
d l^2 &=& \gamma_{i j}dx^{i}dx^{j} \\
\gamma_{i j} &=&  \frac{g_{0 0}}{\mathbf{g}_{0 0}}\left(\mathbf{g}_{i j} - \frac{\mathbf{g}_{i 0}\mathbf{g}_{j 0}}{\mathbf{g}_{0 0}}\right). \nonumber
\end{eqnarray}

To guarantee that (\ref{tri metric}) is definite positive, we need:

\begin{eqnarray}
\label{conds gamma}
\gamma_{11} > 0 \textrm{, }
\left|
  \begin{array}{cc}
   \gamma_{11} & \gamma_{12} \\
   \gamma_{21} & \gamma_{22} \\
  \end{array}
\right| > 0 \textrm{ and }
\left|
  \begin{array}{ccc}
   \gamma_{11} & \gamma_{12} & \gamma_{13} \\
   \gamma_{21} & \gamma_{22} & \gamma_{23} \\
   \gamma_{31} & \gamma_{32} & \gamma_{33} \\
  \end{array}
\right| > 0.
\end{eqnarray}

Therefore, we measure proper time using the metric $g_{\mu \nu}$, but the space geometry is determined by both tensor fields, $g_{\mu \nu}$ and $\tilde{g}_{\mu \nu}$. In the next section, we will study the Schwarzschild Case to apply it in gravitational lensing, perihelion precession and briefly to black holes.\\[30pt]

\newpage

\section{\label{Sec: Schwarzschild Case}Schwarzschild Case.}

The initial foundation of $\tilde{\delta}$ Gravity consist in obtaining a quantum gravity model \cite{delta gravity}. Then, we developed a classical analysis in a cosmological level. In first place, the accelerated expansion of the universe was explained without a cosmological constant \cite{DG DE}-\cite{Paper 1} and $\tilde{\delta}$ Matter was studied to explain the Dark Matter phenomenon \cite{Paper 2}. In this paper, we will develop $\tilde{\delta}$ Gravity in a Schwarzschild geometry to study the principal phenomena used to test GR, the deflection of light by gravitational lensing and the perihelion precession. So, in this case:

\begin{eqnarray}
\label{g Schwar}
g_{\mu \nu}dx^{\mu}dx^{\nu} =  - A(r) c^2dt^2 + B(r) dr^2 + r^2\left(d\theta^2 + \sin^2(\theta)d\phi^2\right).
\end{eqnarray}

For $\tilde{g}_{\mu \nu}$, we can use a similar expression:

\begin{eqnarray}
\label{gt Schwar}
\tilde{g}_{\mu \nu}dx^{\mu}dx^{\nu} =  - \tilde{A}(r) c^2dt^2 + \tilde{B}(r) dr^2 + \tilde{F}(r) r^2\left(d\theta^2 + \sin^2(\theta)d\phi^2\right).
\end{eqnarray}

Before to solve $g_{\mu \nu}$ and $\tilde{g}_{\mu \nu}$, using the equations presented in \textbf{Section \ref{Sec: delta Gravity}}, we need to fix the gauge and the correct boundary conditions.\\

\subsection{\label{SubSec: HG Schwarzschild}Schwarzschild Harmonic Gauge:}

We know that the Einstein's equations do not fix all degrees of freedom of $g_{\mu \nu}$. This means that, if $g_{\mu \nu}$ is solution, then exist other solution $g'_{\mu \nu}$ given by a general coordinate transformation $x \rightarrow x'$. We can eliminate these degrees of freedom by adopting some particular coordinate system, fixing the gauge.One particularly convenient gauge is given by the extended harmonic coordinate conditions. It must be extended because we need to consider the $\tilde{g}_{\mu \nu}$'s components. Then, the gauge fixing is given by (For more details, see \cite{Paper 1}):

\begin{eqnarray}
\label{Harmonic gauge}
\Gamma^{\mu} &\equiv& g^{\alpha \beta}\Gamma^{\mu}_{\alpha \beta} = 0 \\
\label{Harmonic gauge tilde}
\tilde{\delta}\left(\Gamma^{\mu}\right) &\equiv& g^{\alpha \beta}\tilde{\delta}\left(\Gamma^{\mu}_{\alpha \beta}\right) - \tilde{g}^{\alpha \beta}\Gamma^{\mu}_{\alpha \beta} = 0,
\end{eqnarray}

where $\tilde{\delta}\left(\Gamma^{\mu}_{\alpha \beta}\right) = \frac{1}{2}g^{\mu \lambda}\left(D_{\beta}\tilde{g}_{\lambda \alpha}+D_{\alpha}\tilde{g}_{\beta \lambda}-D_{\lambda}\tilde{g}_{\alpha \beta}\right)$. To satisfy (\ref{Harmonic gauge}) and (\ref{Harmonic gauge tilde}), we will use the coordinate transformation:

\begin{eqnarray}
X_1 &=& (r-\mu)\sin(\theta)\cos(\phi) \nonumber \\
X_2 &=& (r-\mu)\sin(\theta)\sin(\phi) \nonumber \\
X_3 &=& (r-\mu)\cos(\theta) \nonumber
\end{eqnarray}

on (\ref{g Schwar}) and (\ref{gt Schwar}), where $\mu = GM$. In this coordinate system, we have:

\begin{eqnarray}
\label{g Schwar 0}
g_{\mu \nu}dx^{\mu}dx^{\nu} &=&  - A(r)c^2dt^2 + \left(\frac{r}{r-\mu}\right)^2 d\textbf{X}^2 + \left(\frac{B(r)}{(r-\mu)^2}-\frac{r^2}{(r-\mu)^4}\right)\left(\textbf{X}\cdot d\textbf{X}\right)^2 \\
\label{gt Schwar 0}
\tilde{g}_{\mu \nu}dx^{\mu}dx^{\nu} &=&  - \tilde{A}(r) c^2dt^2 + \tilde{F}(r)\left(\frac{r}{r-\mu}\right)^2 d\textbf{X}^2 + \left(\frac{\tilde{B}(r)}{(r-\mu)^2} - \frac{\tilde{F}(r)r^2}{(r-\mu)^4}\right)\left(\textbf{X}\cdot d\textbf{X}\right)^2,
\end{eqnarray}

where $r = \mu + \sqrt{X_1^2+X_2^2+X_3^2}$. This system is not convenient to work, so we will fix the gauge in the harmonic coordinate and then we will return to the standard coordinate system, given by (\ref{g Schwar}) and (\ref{gt Schwar}). Now, (\ref{g Schwar 0}) satisfy (\ref{Harmonic gauge}) automatically, but (\ref{Harmonic gauge tilde}) say us that (\ref{gt Schwar 0}) needs an additional condition. We will see it below.\\

\subsection{\label{SubSec: Schwarzschild solution}Schwarzschild solution:}

The correct boundary conditions, that give us the correct Minkowski limit, are $g_{\mu \nu} \rightarrow \eta_{\mu \nu}$ and $\tilde{g}_{\mu \nu} \rightarrow 0$ for $r \rightarrow \infty$\footnote{In \cite{delta gravity}, we suggested that the boundary condition is given by $\tilde{g}_{\mu \nu} \rightarrow \eta_{\mu \nu}$ for $r \rightarrow \infty$ (see eq. (48) in the reference), but recently we noticed that the conditions presented in this paper are the correct choice. That is because $g_{\mu \nu} \rightarrow \eta_{\mu \nu}$, $\tilde{g}_{\mu \nu} \equiv \tilde{\delta}g_{\mu \nu}$ and $\tilde{\delta}\eta_{\mu \nu} = 0$, so it is natural to use $\tilde{g}_{\mu \nu} \rightarrow 0$ for $r \rightarrow \infty$.}. Now, we can solve the equations of motion for the Schwarzschild metric. To simplify the problem, we will solve the equations in empty space, this means the region where $\tilde{T}_{\mu \nu} = T_{\mu \nu} = 0$. The solutions of our equations of motion (\ref{Einst Eq}) and (\ref{tilde Eq}) are:

\begin{eqnarray}
\label{Schwar Sol 1}
A(r) &=& 1 - \frac{2\mu}{r} \\
\label{Schwar Sol 2}
B(r) &=& \frac{1}{1 - \frac{2\mu}{r}} \\
\label{Schwar Sol 3}
\tilde{B}(r) &=& \frac{r^2(r-2\mu)\tilde{A}'(r) - 2\mu r \tilde{A}(r) + r(r-2\mu)(r-\mu)\tilde{F}'(r) + r(r-2\mu)\tilde{F}(r)}{(r-2\mu)^2}
\end{eqnarray}

and survive the equation:

\begin{eqnarray}
\label{Schwar EQ 1}
r\tilde{A}''(r) + 2\tilde{A}'(r) - \mu \tilde{F}''(r) = 0,
\end{eqnarray}

where $' = \frac{d}{dr}$. (\ref{Schwar Sol 1}) and (\ref{Schwar Sol 2}) is the well-known Schwarzschild solution to Einstein equations, where we imposed $A(\infty) = B(\infty) = 1$, to obtain $g_{\mu \nu} \rightarrow \eta_{\mu \nu}$, when $r \rightarrow \infty$. As we said in \textbf{Section \ref{SubSec: HG Schwarzschild}}, we need to fix the gauge for $\tilde{g}_{\mu \nu}$ to obtain an additional equation of $\tilde{A}(r)$ and $\tilde{F}(r)$. This equation comes from (\ref{Harmonic gauge tilde}), so:

\begin{eqnarray}
\label{Schwar gauge eq}
r^2(r-2\mu)\tilde{A}''(r) + 4r(r-2\mu)\tilde{A}'(r) - 4\mu \tilde{A}(r) + r(r-2\mu)(r-\mu)\tilde{F}''(r) + 4(r-\mu)^2\tilde{F}'(r) = 0.
\end{eqnarray}

Therefore, the general solution of (\ref{Schwar EQ 1}) and (\ref{Schwar gauge eq}) is given by:

\begin{eqnarray}
\label{tF}
\tilde{F}(r) &=& \tilde{F}_1 - \int_{\infty}^r \frac{u^3\left(u-2\mu\right)\tilde{A}''(u) + 2u\left(u+\mu\right)\left(u-2\mu\right)\tilde{A}'(u) - 4\mu\tilde{A}(u)}{4\mu\left(u-\mu\right)^2} du \\
\tilde{A}(r) &=& \frac{\tilde{A}_1\mu(r-\mu)}{r^2} + \frac{\tilde{A}_2(r^2-2\mu^2)}{r^2} + \tilde{A}_3\mu\frac{2\mu + (r-\mu)\ln\left(1-\frac{2\mu}{r}\right)}{r^2},
\end{eqnarray}

where $\tilde{A}_1$, $\tilde{A}_2$ and $\tilde{A}_3$ are integration constants. By the boundary conditions, we have to impose the conditions $\tilde{A}(\infty) = \tilde{B}(\infty) = \tilde{F}(\infty) = 0$ to obtain $\tilde{g}_{\mu \nu} \rightarrow 0$, when $r \rightarrow \infty$. These conditions just means that $\tilde{A}_2 = 0$ and $\tilde{F}_1 = 0$. Then, the solutions are ($\tilde{A}_1 = - 2a_0$ and $\tilde{A}_3 = - a_1$):

\begin{eqnarray}
\label{Schwar Sol 4}
\tilde{A}(r) &=& - \frac{2a_0\mu(r-\mu)}{r^2} - a_1\mu\frac{2\mu + (r-\mu)\ln\left(1-\frac{2\mu}{r}\right)}{r^2} \\
\label{Schwar Sol 5}
\tilde{F}(r) &=& \frac{2a_0\mu}{r} - a_1\frac{2\mu + (r-\mu)\ln\left(1-\frac{2\mu}{r}\right)}{r} \\
\label{Schwar Sol 6}
\tilde{B}(r) &=& \frac{2a_0\mu(r-\mu)}{(r-2\mu)^2} - a_1\frac{2\mu(r-2\mu) + (r^2-3\mu r+\mu^2)\ln\left(1-\frac{2\mu}{r}\right)}{(r-2\mu)^2}.
\end{eqnarray}

So, we have three parameters: $\mu$ comes from the ordinary metric components and represent the mass of a massive object (planets, stars, black holes, etc). Finally, we have $a_0$ and $a_1$. These parameters are adimensional and represent the correction by $\tilde{\delta}$ Gravity. Later, we will understand the physical meaning of these parameters.\\

Now, we must remember that (\ref{Schwar Sol 1}), (\ref{Schwar Sol 2} and (\ref{Schwar Sol 4}-\ref{Schwar Sol 6}) correspond to the solution in the region without matter. This means that $r > R$, where $R$ is the radius of a star for example. Generally, the Newtonian approximation can be used, so that $R \gg 2\mu$. However the logarithmic solution could be important in black holes, where the Newtonian approximation is not valid. So, considering the leading order in $\frac{\mu}{r}$, the solution is reduced to:

\begin{eqnarray}
\label{Schwar Sol 1 aprox}
A(r) &=& 1 - \frac{2\mu}{r} \\
\label{Schwar Sol 2 aprox}
B(r) &=& 1 + \frac{2\mu}{r} + O\left(\left(\frac{2\mu}{r}\right)^2\right) \\
\label{Schwar Sol 4 aprox}
\tilde{A}(r) &=& - \frac{2a_0\mu}{r} + O\left(\left(\frac{2\mu}{r}\right)^2\right) \\
\label{Schwar Sol 5 aprox}
\tilde{F}(r) &=& \frac{2a_0\mu}{r} + O\left(\left(\frac{2\mu}{r}\right)^2\right) \\
\label{Schwar Sol 6 aprox}
\tilde{B}(r) &=& \frac{2a_0\mu}{r} + O\left(\left(\frac{2\mu}{r}\right)^2\right).
\end{eqnarray}

Notice that $a_1$ disappears in (\ref{Schwar Sol 1 aprox}-\ref{Schwar Sol 6 aprox}). This means that this parameter is only important in a Post-Newtonian approximation. We will use these expressions in the next section to describe the Gravitational Lensing effect.\\

\subsection{\label{SubSec: Gravitational Lensing}Gravitational Lensing:}

\begin{figure}
\begin{center}
\includegraphics[scale=0.4]{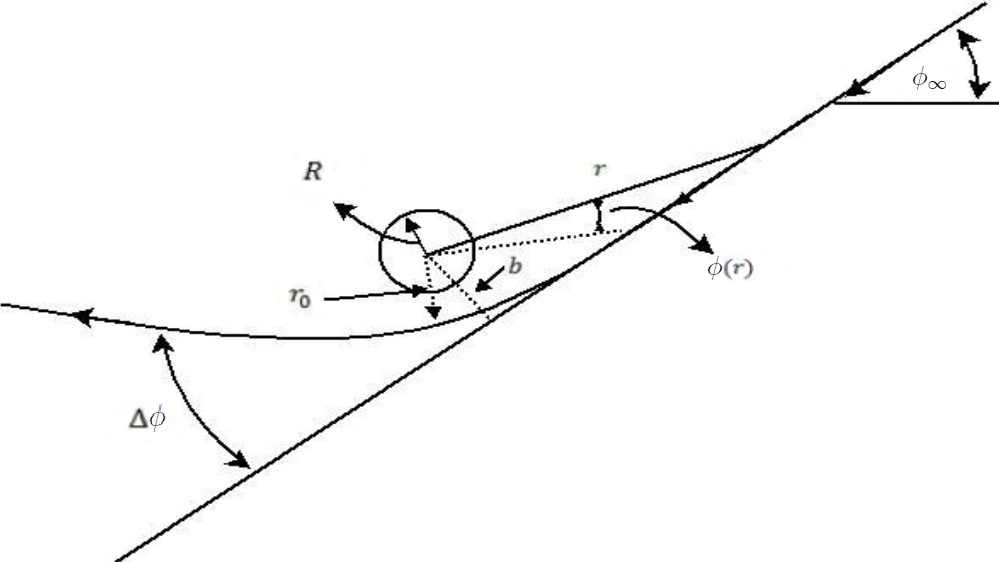}
\caption{{\scriptsize Trajectory by gravitational lensing. $R$ is the radius of the star, $r_0$ is the minimal distance to the star, $b$ is the impact parameter, $\phi_{\infty}$ is the incident direction and $\Delta \phi$ is the deflection of light.}}
\label{Fig: Grav Lens}
\end{center}
\end{figure}

To describe this phenomenon, we need the null geodesic, in our case, given by (\ref{geodesics null}). To solve these equations, we will consider a coordinate system where $\theta = \frac{\pi}{2}$ such that the trajectory is given by \textbf{Figure \ref{Fig: Grav Lens}} (See for instance \cite{Weinberg grav}).\\

The geodesic equations are complicated, but with some work, we may reduce it to:

\begin{eqnarray}
\label{Schwar v0}
\frac{d t}{d u} &=& \frac{E}{A(r)+\tilde{A}(r)} \\
\label{Schwar vr}
\left(\frac{d r}{d u}\right)^2 &=& \frac{E^2}{\left(A(r)+\tilde{A}(r)\right)\left(B(r)+\tilde{B}(r)\right)} - \frac{J^2}{r^2\left(B(r)+\tilde{B}(r)\right)\left(1+\tilde{F}(r)\right)} \\
\label{Schwar vphi}
\frac{d \phi}{d u} &=& \frac{J}{r^2\left(1+\tilde{F}(r)\right)},
\end{eqnarray}

where $u$ is the trajectory parameter such that $x^{\mu} = x^{\mu}(u)$ and $E$ and $J$ are constants of motion. From (\ref{Schwar v0}), we can see that $t \rightarrow Eu$ when $r \rightarrow \infty$. From (\ref{Schwar vr}), we can define a \textit{effective potential} given by:

\begin{eqnarray}
\label{V eff}
V_{eff}(r) = 1 - \frac{1}{\left(A(r)+\tilde{A}(r)\right)\left(B(r)+\tilde{B}(r)\right)} + \frac{\left(J/E\right)^2}{r^2\left(B(r)+\tilde{B}(r)\right)\left(1+\tilde{F}(r)\right)},
\end{eqnarray}

such that:

\begin{eqnarray}
v_r^2(r) \equiv \frac{1}{E^2}\left(\frac{d r}{d u}\right)^2 = 1 - V_{eff}(r).
\end{eqnarray}

Using (\ref{Schwar Sol 1}-\ref{Schwar Sol 3}) and (\ref{Schwar Sol 4}-\ref{Schwar Sol 6}), we can see that $V_{eff}(\infty) = 0$ and $v_r(\infty) = 1$. To obtain a plot of $V_{eff}(r)$, we need to fix $a_0$ and $a_1$.\\

Finally, from \textbf{Figure \ref{Fig: Grav Lens}}, we know that $\phi(r)$ is necessary to study the gravitational lensing. For this, we use (\ref{Schwar vr}) and (\ref{Schwar vphi}) to obtain:

\begin{eqnarray}
\label{Schwar dphi(r)}
\left(r^2\left(\frac{d \phi}{d r}(r)\right)\right)^{-2} &=& \frac{1+\tilde{F}(r)}{B(r)+\tilde{B}(r)}\left(\frac{E^2\left(1+\tilde{F}(r)\right)}{J^2\left(A(r)+\tilde{A}(r)\right)} - \frac{1}{r^2}\right) \nonumber \\
\left(y^2\left(\frac{d \phi}{d y}(y)\right)\right)^{-2} &=& \frac{1+\tilde{F}(r_0y)}{B(r_0y)+\tilde{B}(r_0y)}\left(\frac{\left(1+\tilde{F}(r_0y)\right)\left(A(r_0)+\tilde{A}(r_0)\right)}{\left(1+\tilde{F}(r_0)\right)\left(A(r_0y)+\tilde{A}(r_0y)\right)} - \frac{1}{y^2}\right),
\end{eqnarray}

where $y = \frac{r}{r_0} \geq 1$ is a normalized radius with $r_0$ the minimal radius, given by $\frac{d r}{d u}|_{r=r_0} = 0$. So:

\begin{eqnarray}
\left(\frac{J}{E}\right)^2 &=& r_0^2\frac{1+\tilde{F}(r_0)}{A(r_0)+\tilde{A}(r_0)}. \nonumber
\end{eqnarray}

Then, the deflection of light can be obtained solving (\ref{Schwar dphi(r)}). However, the approximation $r_0 >> 2\mu$ is usually used to obtain a explicit result of $\Delta \phi$ when the matter source is not dense enough. So, with this approximation, we obtain:

\begin{eqnarray}
\label{Schwar dphi(r) aprox 0}
\left(\frac{d \phi}{d y}(y)\right)^{-2} &\simeq& y^2\left(y^2-1\right)\left(1-\frac{\epsilon}{y}-\frac{\left(1+2a_0\right)y\epsilon}{1+y}\right) + a_0O\left(\epsilon^2\right),
\end{eqnarray}

where $\epsilon = \frac{2\mu}{r_0}$. We notice that (\ref{Schwar dphi(r) aprox 0}) to first order on $\epsilon$ is exact in GR, but in $\tilde{\delta}$ Gravity we have higher orders terms. Now, to obtain the deflection of light, we develop (\ref{Schwar dphi(r) aprox 0}) such that:

\begin{eqnarray}
\label{Schwar phi(r) aprox}
\frac{d \phi}{d y}(y) &\simeq& \pm \frac{1}{y\sqrt{y^2-1}}\left(1+\frac{\epsilon}{2y}+\frac{\left(1+2a_0\right)y\epsilon}{2\left(1+y\right)}\right) + O\left(\epsilon^2\right) \nonumber \\
\phi(y) - \phi_{\infty} &\simeq& \pm \int_{y}^{\infty} \frac{dy'}{y'\sqrt{y'^2-1}}\left(1+\frac{\epsilon}{2y'}+\frac{\left(1+2a_0\right)y'\epsilon}{2\left(1+y'\right)}\right) + O\left(\epsilon^2\right).
\end{eqnarray}

We want to describe a complete trajectory, so the photon start from $\phi(y = \infty)$ up to $\phi(y = 1)$ and then back again to $\phi(y = \infty)$ (See \textbf{Figure \ref{Fig: Grav Lens}}). Besides, if the trajectory were a straight line, this would equal just $\pi$. All of this means that the deflection of light is:

\begin{eqnarray}
\label{Deflect}
\Delta \phi &=& 2 |\phi(1) - \phi_{\infty}| - \pi \nonumber \\
&\simeq& 2\left|\int_{1}^{\infty} \frac{dy'}{y'\sqrt{y'^2-1}}\left(1+\frac{\epsilon}{2y'}+\frac{\left(1+2a_0\right)y'\epsilon}{2\left(1+y'\right)}\right)\right| - \pi + O\left(\epsilon^2\right) \nonumber \\
&\simeq& 2\left(1+a_0\right)\epsilon + O\left(\epsilon^2\right) \nonumber \\
&\simeq& \frac{4\mu\left(1+a_0\right)}{r_0} + O\left(\left(\frac{2\mu}{r_0}\right)^2\right).
\end{eqnarray}

From GR, $\Delta \phi = \frac{4\mu}{r_0}$. So, in our modified gravity, we have an additional term given by $\frac{4\mu a_0}{r_0}$. On the other side, we have an experimental value $\Delta \phi_{Exp} = 1.761'' \pm 0.016''$ for the sun \cite{Deflect} and it is very close to the prediction of GR. This means that, to satisfy the experimental value with $\tilde{\delta}$ Gravity, it is necessary that our additional term provide a very small correction, such that:

\begin{eqnarray}
\label{Schwar result}
\left|\frac{4\mu a_0}{r_0}\right| &=& 1.761''|a_0| < 0.016'' \nonumber \\
|a_0| &<& 0.009086.
\end{eqnarray}

From (\ref{Deflect}), we can see that $a_0$ represents an additional mass by $\tilde{\delta}$ Matter given by $M_{add} = a_0 M$, where $M$ is the solar mass. So, the result of (\ref{Schwar result}) tells us that $\tilde{\delta}$ Matter must be less than $1 \%$ close to the Sun. In \cite{dark matter prop} was estimated observationally that the Dark Matter mass in the sphere within Saturn's orbit should be less than $1.7 \times 10^{-10} \%$. On the other side, we expect that $\tilde{\delta}$ Matter will be more important in a galactic scale. Then, Dark Matter could be explained with $\tilde{\delta}$ Matter. An analysis developed in this way is presented in \cite{Paper 2}.\\

\subsection{\label{SubSec: Perihelion Precession}Perihelion Precession:}

In the last section, we used the null geodesic to compute the deflection of light. However, if we want to study the trajectory of a massive object, we need (\ref{geodesics m}). These equations, for $\theta = \frac{\pi}{2}$, are given by (See for instance \cite{Weinberg grav}):

\begin{eqnarray}
\label{Schwar vr mass}
\left(\frac{d r}{d u}\right)^2 &=& \frac{1}{B(r)}\left(\left(A(r) - \frac{J^2}{r^2\left(\frac{E}{A(r)}+\left(\tilde{F}(r)-\frac{\tilde{A}(r)}{A(r)}\right)\left(\frac{d t}{d u}\right)\right)^2}\right)\left(\frac{d t}{d u}\right)^2 - 1\right) \\
\label{Schwar vphi mass}
\frac{d \phi}{d u} &=& \frac{J\left(\frac{d t}{d u}\right)}{r^2\left(\frac{E}{A(r)}+\left(\tilde{F}(r)-\frac{\tilde{A}(r)}{A(r)}\right)\left(\frac{d t}{d u}\right)\right)},
\end{eqnarray}

where $\frac{d t}{d u}$ obey a fifth order equation:

\begin{eqnarray}
\label{Schwar v0 mass}
\left(1 + \frac{1}{2}\left(\frac{2\tilde{A}(r)}{A(r)}-\frac{\tilde{B}(r)}{B(r)}+\left(A(r)\left(\frac{\tilde{B}(r)}{B(r)}-\frac{\tilde{A}(r)}{A(r)}\right) + \frac{J^2\left(\tilde{F}(r)-\frac{\tilde{B}(r)}{B(r)}\right)}{r^2\left(\frac{E}{A(r)}+\left(\tilde{F}(r)-\frac{\tilde{A}(r)}{A(r)}\right)\left(\frac{d t}{d u}\right)\right)^2}\right)\left(\frac{d t}{d u}\right)^2\right)\right)\left(\frac{d t}{d u}\right) = \frac{E}{A(r)}.
\end{eqnarray}

These equations are very difficult to solve, but in the approximation $r_0 >> 2\mu$ with $r_0$ the minimal radius, we have:

\begin{eqnarray}
\label{Schwar v0 mass approx}
\frac{1}{E}\left(\frac{d t}{d u}\right) &\simeq& 1 + \left(1-a_0\left(j^2-\frac{1}{2}\right)\right)\frac{\epsilon}{y} \\
&& + \left(1+\frac{a_0}{4y^2}\left(j^2\left(4y^3-10y^2+1\right)+y^2\left(4y-1\right)\right)+\frac{a_0^2}{4}\left(2j^2+1\right)\left(4y+6j^2-3\right)\right)\frac{\epsilon^2}{y^2} + O\left(\epsilon^3\right) \nonumber \\
\label{Schwar vr mass approx}
\left(\frac{d r}{d u}\right)^2 &\simeq& j^2\left(1-\frac{1}{y^2}\right)\left[1-\frac{1+a_0\left(2j^2+3\right)}{y}\epsilon - \frac{y\left(1+j^2+a_0\left(1+2j^2\right)\right)}{j^2\left(1+y\right)}\epsilon \right. \nonumber \\
&& + \frac{a_0}{4}\left(\frac{2j^2}{y^4}-\frac{4j^2-a_0\left(28j^4+44j^2+19\right)}{y^2}+\frac{8\left(1+j^2+a_0\left(2j^2+1\right)\right)}{y} \right. \nonumber \\
&& \left. \left. + \frac{2+6j^2+a_0\left(1+12j^2+4j^4\right)}{j^2} + \frac{4\left(1+3j^2+2j^4+a_0\left(1+2j^2\right)^2\right)}{j^2\left(1+y\right)}\right)\epsilon^2 + O\left(\epsilon^3\right)\right] \\
\label{Schwar vphi mass approx}
\frac{d \phi}{d u} &\simeq& \frac{j}{r_0y^2}\Biggl[1-a_0\left(j^2+\frac{3}{2}\right)\frac{\epsilon}{y} \\
&& \left. - a_0\left(\frac{5}{4}-y-\frac{j^2\left(2y-1\right)\left(2y^2-2y-1\right)}{4y^2}-a_0\left(3j^4+2j^2\left(2+y\right)+y+\frac{5}{4}\right)\right)\frac{\epsilon^2}{y^2} + O\left(\epsilon^3\right)\right], \nonumber
\end{eqnarray}

where $y = \frac{r}{r_0}$, $\epsilon = \frac{2\mu}{r_0}$, $j = \frac{J}{r_0}$ and $\frac{d r}{d u}|_{r=r_0} = 0$, so:

\begin{eqnarray}
E^2 \simeq 1 + j^2 - \left(1+j^2+a_0\left(2j^2+1\right)\right)\epsilon + \frac{a_0}{2}\left(1+3j^2+a_0\left(2j^4+6j^2+\frac{1}{2}\right)\right)\epsilon^2 + O(\epsilon^3). \nonumber
\end{eqnarray}

From (\ref{Schwar vr mass approx}-\ref{Schwar vphi mass approx}), we obtain:

\begin{eqnarray}
\label{diff phi(r) mass}
\left(\frac{d \phi}{d y}(y)\right)^2 &\simeq& y^{-2}\left(y^2-1\right)^{-1}\left[1 + \left(1 + \frac{y^2\left(1+j^2+a_0\left(2j^2+1\right)\right)}{j^2\left(1+y\right)}\right)\frac{\epsilon}{y} \right. \nonumber \\
&& + \left(\frac{1+\frac{a_0}{2}}{y^2}+\frac{\left(1+j^2+a_0\left(2j^2+1\right)\right)^2}{j^4\left(y+1\right)^2}-\frac{2\left(1+a_0\left(j^2+1\right)\right)\left(1+j^2+a_0\left(2j^2+1\right)\right)}{j^4\left(y+1\right)} \right. \nonumber \\
&& \left. \left. + \frac{4\left(j^2+1\right)^2+2a_0\left(5j^4+11j^2+4\right)-a_0^2\left(4j^6-4j^4-15j^2-4\right)}{4j^4}\right)\epsilon^2 + O\left(\epsilon^3\right)\right].
\end{eqnarray}

However, it is useful to rewrite (\ref{diff phi(r) mass}) using:

\begin{eqnarray}
\lambda = \frac{\lambda_0}{y}, \textrm{ with } \lambda_0 \simeq \frac{2j^2}{\epsilon\left(1+a_0\left(2j^2+1\right)-2a_0\left(1+j^2+a_0\left(2j^2+1\right)\right)\epsilon\right)}. \nonumber
\end{eqnarray}

That is:

\begin{eqnarray}
\label{diff lambda(phi) mass}
\left(\frac{d \lambda}{d \phi}(\phi)\right)^2 \simeq \bar{e} - 1 + 2\lambda(\phi) - \left(1-\frac{a_0\left(6-a_0\left(4j^4-4j^2-7\right)\right)\varepsilon}{2\left(1+a_0\left(2j^2+1\right)\right)}\right)\lambda^2(\phi) + \varepsilon \lambda^3(\phi) + \frac{a_0\varepsilon^2}{2} \lambda^4(\phi) + a_0O\left(\varepsilon^3\right),
\end{eqnarray}

where:

\begin{eqnarray}
\label{New Epsilon}
\varepsilon &\simeq& \frac{\left[1+a_0\left(2j^2+1\right)-2a_0\left(1+j^2+a_0\left(2j^2+1\right)\right)\epsilon\right]\epsilon^2}{2j^2} \\
\label{Energy Redef}
\bar{e} &\simeq& 1 + \frac{j^2\left[4j^2-4\left(1+j^2+a_0\left(2j^2+1\right)\right)\epsilon+a_0\left(2+6j^2+a_0\left(4j^4+12j^2+1\right)\right)\epsilon^2\right]}{\left(1+a_0\left(2j^2+1\right)
-2a_0\left(1+j^2+a_0\left(2j^2+1\right)\right)\epsilon\right)^2\epsilon^2} \nonumber \\
&\simeq& 1 + \frac{2\left(E^2-1\right)}{\left(1+a_0\left(2j^2+1\right)-2a_0\left(1+j^2+a_0\left(2j^2+1\right)\right)\epsilon\right)\varepsilon},
\end{eqnarray}

with $\bar{e}$ the orbital eccentricity. In GR, (\ref{diff lambda(phi) mass}) is exact to first order on $\varepsilon$ and the cubic term in $\lambda(\phi)$ explain the mercury's perihelion precession. In $\tilde{\delta}$ Gravity we have high order corrections too, but they are practically suppressed when $r_0 >> 2\mu$. Besides, (\ref{Energy Redef}) can be interpreted as an energy redefinition, such that the \textit{new energy} is given by $\tilde{E}^2 \sim 1 + \frac{\left(E^2-1\right)}{\left(1+a_0\left(2j^2+1\right)\right)}$, but this modification must be small to satisfy (\ref{Schwar result}), so the orbital movements are equal to GR. On the other side, other corrections appear when $r_0$ is smaller, close to $2\mu$. In that case, we must use the exact equations, given by (\ref{Schwar vr mass}-\ref{Schwar v0 mass}), but they are very complicated. However, we can try to solve them in a particular radius, for example $r_0$ where $\frac{dr}{du} = 0$. In that case, (\ref{Schwar vr mass}) and (\ref{Schwar v0 mass}) can be reduced to:

\begin{eqnarray}
\label{EQ r0 1}
\left(A(r_0) - \frac{j^2}{\left(\frac{E}{A(r_0)}+\left(\tilde{F}(r_0)-\frac{\tilde{A}(r_0)}{A(r_0)}\right)\left.\left(\frac{d t}{d u}\right)\right|_{r_0}\right)^2}\right)\left.\left(\frac{d t}{d u}\right)^2\right|_{r_0} &=& 1 \\
\label{EQ r0 2}
\left(1 + \frac{\tilde{A}(r_0)}{A(r_0)} - \frac{\tilde{F}(r_0)}{2} + \frac{A(r_0)}{2}\left(\tilde{F}(r_0)-\frac{\tilde{A}(r_0)}{A(r_0)}\right)\left.\left(\frac{d t}{d u}\right)^2\right|_{r_0}\right)\left.\left(\frac{d t}{d u}\right)\right|_{r_0} &=& \frac{E}{A(r_0)},
\end{eqnarray}

where $j = \frac{J}{r_0}$. They are a fourth and a third order equations in $\left.\left(\frac{d t}{d u}\right)\right|_{r_0}$ respectively. So, by iteration, we can reduce the order of these equation, such as:

\begin{eqnarray}
\label{dtdu}
\left.\left(\frac{d t}{d u}\right)\right|_{r_0} = \frac{E}{A(r_0)}\Sigma(r_0),
\end{eqnarray}

with:

\begin{eqnarray}
\label{EQ r0 1 sol}
\Sigma(r_0) &=& \frac{1 - f_1(r_0)f_2(r_0)}{\left(1 + \frac{\tilde{A}(r_0)}{A(r_0)} - \frac{\tilde{F}(r_0)}{2}\right)\left(1 - \frac{4}{3}f_1(r_0)f_2(r_0)\right) + f_1(r_0)} \\
f_1(r_0) &=& \frac{3E^2\left(\frac{\tilde{A}(r_0)}{A(r_0)}-\tilde{F}(r_0)\right)}{2A(r_0)\left(\frac{E^2}{A(r_0)}-j^2+\left(2+\frac{\tilde{A}(r_0)}{A(r_0)}\right)\left(\frac{\tilde{A}(r_0)}{A(r_0)}-\tilde{F}(r_0)\right)\right)} \nonumber \\
f_2(r_0) &=& \frac{4\left(1 + \frac{\tilde{A}(r_0)}{A(r_0)} - \frac{\tilde{F}(r_0)}{2}\right)}{\frac{E^2}{A(r_0)}-j^2+\left(2+\frac{\tilde{A}(r_0)}{A(r_0)}\right)\left(\frac{\tilde{A}(r_0)}{A(r_0)}-\tilde{F}(r_0)\right)} \nonumber
\end{eqnarray}

and the minimum radius can be obtained numerically from:

\begin{eqnarray}
\label{EQ r0 2 sol}
\left(\frac{E^2}{A(r_0)}-j^2+\left(2+\frac{\tilde{A}(r_0)}{A(r_0)}\right)\left(\frac{\tilde{A}(r_0)}{A(r_0)}-\tilde{F}(r_0)\right)\right)\Sigma^2(r_0) - 4\left(1 + \frac{\tilde{A}(r_0)}{A(r_0)} - \frac{\tilde{F}(r_0)}{2}\right)\Sigma(r_0) + 3 = 0,
\end{eqnarray}

where we have to use the exact solution of $A(r_0)$, $B(r_0)$, $\tilde{A}(r_0)$, $\tilde{B}(r_0)$ and $\tilde{F}(r_0)$ in \textbf{Section \ref{SubSec: Schwarzschild solution}}. In the process, we imposed that (\ref{EQ r0 1}) and (\ref{EQ r0 2}) possess one pole in common for $\Sigma(r_0)$. $r_0$ is an important element to understand the orbital trajectories, but we need to fix $a_0$ and $a_1$ to solve it. On the other side, we can verify that in the limit where $\left(\tilde{A}(r_0),\tilde{B}(r_0),\tilde{F}(r_0)\right) \rightarrow 0$, our results are reduced to GR. That is:

\begin{eqnarray}
(\ref{EQ r0 1 sol}) &\rightarrow& \left.\left(\frac{d t}{d u}\right)\right|_{r_0} = \frac{E}{A(r_0)} = \frac{E}{1-\frac{2\mu}{r_0}} \nonumber \\
(\ref{EQ r0 2 sol}) &\rightarrow&  E^2 = A(r_0)\left(1+j^2\right) \rightarrow r_0 = \frac{2\mu\left(1+j^2\right)}{1+j^2-E^2}. \nonumber
\end{eqnarray}

In conclusion, in the last two sections we saw that $\tilde{\delta}$ Gravity give us important corrections to orbital trajectories, but they do not produce big differences with GR when the trajectory is far away from the Schwarzschild radius, $r_s = 2\mu$, on the condition that $a_0$ is small enough (See equation (\ref{Schwar result})). This is always true for stars, planets and any low-density object. We saw that $a_0$ represents the $\tilde{\delta}$ Matter contribution, however the physical meaning of $a_1$ is unknown yet. To solve this, the study of massive object is necessary. To finish, in the next section, we will introduce the analysis of Black Holes to complete the Schwarzschild case.\\

\subsection{\label{SubSec: Black Holes}Black Holes:}

In \textbf{Section \ref{SubSec: Test Particle}}, we saw that the proper time is defined using the metric $g_{\mu \nu}$, such as GR where $g_{\mu \nu}\dot{x}^{\mu}\dot{x}^{\nu} = -1$. Then, the proper time is given by (\ref{proper time}). On the other side, the null geodesic of massless particles say us the the space geometry is determined by both tensor fields, $g_{\mu \nu}$ and $\tilde{g}_{\mu \nu}$. This means that the three-dimensional metric is given by (\ref{tri metric}). Then, for a Schwarzschild geometry, it is reduced to:

\begin{eqnarray}
\label{tri metric Schw}
d l^2 &=& \frac{A(r)\left(B(r)+\tilde{B}(r)\right)}{\left(A(r)+\tilde{A}(r)\right)} dr^2 + \frac{r^2A(r)\left(1+\tilde{F}(r)\right)}{\left(A(r)+\tilde{A}(r)\right)}\left(d\theta^2 + \sin^2(\theta)d\phi^2\right).
\end{eqnarray}

Now, if we apply the conditions (\ref{conds gamma}) plus $g_{00} < 0$ to (\ref{tri metric Schw}), we obtain:

\begin{eqnarray}
\label{rules}
A(r) > 0 \wedge f_A(r) > 0 \wedge f_B(r) > 0 \wedge f_F(r) > 0,
\end{eqnarray}

where:

\begin{eqnarray}
\label{fA}
f_A(r) &=& A(r) + \tilde{A}(r) \nonumber \\
&=& 1 - \frac{2\mu}{r} - \frac{2a_0\mu\left(r-\mu\right)}{r^2} - a_1\mu\frac{2\mu + \left(r-\mu\right)\ln\left(1-\frac{2\mu}{r}\right)}{r^2} \\
\label{fB}
f_B(r) &=& B(r) + \tilde{B}(r) \nonumber \\
&=& \frac{r}{r-2\mu} + \frac{2a_0\mu\left(r-\mu\right)}{\left(r-2\mu\right)^2} - a_1\frac{2\mu\left(r-2\mu\right) + \left(r^2-3\mu r+\mu^2\right)\ln\left(1-\frac{2\mu}{r}\right)}{\left(r-2\mu\right)^2} \\
\label{fF}
f_F(r) &=& 1 + \tilde{F}(r) \nonumber \\
&=& 1 + \frac{2a_0\mu}{r} - a_1\frac{2\mu + \left(r-\mu\right)\ln\left(1-\frac{2\mu}{r}\right)}{r}.
\end{eqnarray}

We can see that these rules are automatically satisfied when $r \gg 2\mu$, so we must consider extreme cases to prove these conditions. In GR, they are reduced to $A(r) > 0$ and $B(r) > 0$, then $r > 2\mu$. This means that these rules define the black hole horizon, $r_H = 2\mu$. Therefore, (\ref{rules}) defines a modified horizon to $\tilde{\delta}$ Gravity, but we need to fix $a_0$ and $a_1$ first.\\

Motivated by the result in \textbf{Section \ref{SubSec: Gravitational Lensing}}, we know that $a_0$ is related to $\tilde{\delta}$ Matter, so $a_0>0$ and probably bigger than (\ref{Schwar result}) to consider a higher quantity of $\tilde{\delta}$ Matter (Dark Matter). Just as an example, we can choose some combination:

\begin{itemize}
  \item If $a_0 = 1$ and $a_1 = 1$, we have $r_H = 3.37\mu$.
  \item If $a_0 = 1$ and $a_1 = -1$, we have $r_H = 3.46\mu$.
\end{itemize}

In both cases, the horizon radius is given by $f_A(r_H) = 0$. For example, if $\left|a_1\right| \gg 1$ and $a_0 \sim 1$, then the horizon is given by $f_B(r)$ ($a_1 > 0$) or $f_F(r)$ ($a_1 < 0$). In any case, we will obtain a horizon radius $\geq r_H = 2\mu$.\\

In GR, the event horizon radius is defined when $g_{rr}$ component of the metric is null. In fact, when we include Electric Charge and/or Angular Momentum in a black hole, a inner and outer event horizon are produced and, additionally, an ergosphere appears, given by $g_{tt} = 0$, defining different regions around the black hole. In $\tilde{\delta}$ Gravity, we saw that the three-dimensional metric gives us the event horizon radiuses and, in the same way than GR, different regions are produced, but Electric Charge or Angular Momentum are not necessary. In a Schwarzschild black hole, these regions are produced whenever the conditions in (\ref{rules}) are violated. Now, the nature of these regions will depend of the value of $a_0$ and $a_1$. Finally, the minimum radius of the orbital trajectory of a massive particle is given by (\ref{EQ r0 2 sol}). All These will developed in a future work.\\

In conclusion, we have that the effects of $\tilde{\delta}$ Gravity are represented by $a_0$ and $a_1$. $a_0$ has a clear physical meaning, it represents the quantity of $\tilde{\delta}$ Matter and could be considered as Dark Matter \cite{Paper 2}. On the other side, the meaning of $a_1$ is more difficult to define. This parameter only appears when we consider a highly massive object, as black holes, defining different kind of regions around of this object. In that sense, it could redefine the concept of black holes. For that reason, black holes in $\tilde{\delta}$ Gravity must be studied in more detail.\\

\newpage

\section*{Conclusions.}

A modified model of gravity called $\tilde{\delta}$ Gravity is presented, where a new gravitational field, $\tilde{g}_{\mu \nu}$, is introduced. Additionally, a new kind of matter fields are included, called $\tilde{\delta}$ Matter fields. All these new fields exhibit a new symmetry: the $\tilde{\delta}$ symmetry, such as the new action is invariant under these transformations. The quantum analysis was developed in \cite{delta gravity}.\\

In previous works, we have studied, in a classical level, the cosmological case in $\tilde{\delta}$ Gravity. We know that the Einstein's equation are preserved and obtain additional equations to solve $\tilde{\delta}$ component. Besides, we have that the free trajectory of a massive particle is given by an anomalous geodesic presented in (\ref{geodesics m}) and massless particles move in a effective null geodesic given by $\mathbf{g}_{\mu \nu} = g_{\mu \nu} + \tilde{g}_{\mu \nu}$. With all these, we explained the accelerated expansion of the universe without a cosmological constant \cite{DG DE}-\cite{Paper 1}.\\

Additionally, in \cite{Paper 2}, the Non-Relativistic limit was studied. We found that $\tilde{\delta}$ Matter is related to ordinary matter and it is used to explain Dark Matter in different scales. We obtained that the $\tilde{\delta}$ Matter effect is only important when the distribution of ordinary matter is strongly dynamic, so the scale is not so important. Additionally, $\tilde{\delta}$ Matter produce an amplifying effect in the total mass and rotation velocity in a galaxy. Besides, it has a special behavior, more similar to Dark Matter compared with its equivalent ordinary component. In \cite{Paper 1} is computed the $\tilde{\delta}$ Matter quantity in the present at cosmological level. We obtained that the $\tilde{\delta}$ non-relativistic Matter is 23\% the ordinary non-relativistic matter, where Dark Matter is included, implying that Dark Matter is in part $\tilde{\delta}$ Matter.\\

In this paper, we analyzed the Schwarzschild case outside matter. We found an exact solution for the equations of motion. We used the Newtonian approximation in these solutions to find the deflection of light by the Sun. To explain the experimental data, the correction must be small, such that $\tilde{\delta}$ Matter is $<1 \%$ of the total mass at a solar system scale. Then the modification of $\tilde{\delta}$ Gravity must not be important at a solar system scale, such as it is presented in \cite{Paper 2}. Then, we study the perihelion precession. The exact solution is very complicated, because a fifth order equation must be solved. However, even in the Newtonian approximation, we can see interesting, but really small, corrections. Basically, $\tilde{\delta}$ Gravity does not have important corrections to GR for low-density object like stars or planets. This means that we need to study high-density objects like black holes to observe important effects by $\tilde{\delta}$ Gravity. To have an idea how the trajectory of a massive particle is affected by massive objects, we solved the equations in the minimum radius.\\

Finally, we presented an introduction to Black Holes, where some conditions to guarantee that the three-dimensional metric is definite positive are studied. We know that GR defines the event horizon radius where $g_{rr} = 0$, and a inner and outer event horizon radiuses when Electric Charge and/or Angular Momentum are included. Besides, $g_{tt} = 0$ give us an ergosphere, so different regions around the black hole can be defined. On the other side, in $\tilde{\delta}$ Gravity, the three-dimensional metric gives us the event horizon radiuses and, in the same way than GR, different regions are produced whenever the conditions in (\ref{rules}) are violated, even in a Schwarzschild black hole. We understood that $a_0$ gives us the quantity of $\tilde{\delta}$ Matter (Dark Matter), but the meaning of $a_1$ is more difficult to define. In any case, it is only relevant to highly massive object, so it is important to define these regions. Therefore, we need to study black holes in more detail in $\tilde{\delta}$ Gravity. It will be developed in a future work.\\

At this moment, in different works, we have developed some phenomena (Expansion of the Universe, Dark Matter, the Deflection of Light, etc) and we introduced a preliminary discussion of others (Black Holes and Inflation). Further tests of the model must include the computation of the CMB power spectrum, the evolution and formation of large-scale structure in the universe and a more detailed analysis of Dark Matter. These works are in progress now.\\

\newpage

\section*{Appendix A: $\tilde{\delta}$ Theories.}

In this Appendix, we will define the $\tilde{\delta}$ Theories in general and their properties. For more details, see \cite{delta gravity,AppendixA}.\\

\subsection*{$\tilde{\delta}$ Variation:}

These theories consist in the application of a variation represented by $\tilde{\delta}$. As a variation, it will have all the properties of a usual variation such as:

\begin{eqnarray}
\tilde{\delta}(AB)&=&\tilde{\delta}(A)B+A\tilde{\delta}(B) \nonumber \\
\tilde{\delta}\delta A &=&\delta\tilde{\delta}A \nonumber \\
\tilde{\delta}(\Phi_{, \mu})&=&(\tilde{\delta}\Phi)_{, \mu},
\end{eqnarray}

where $\delta$ is another variation. The particular point with this variation is that, when we apply it on a field (function, tensor, etc.), it will give new elements that we define as $\tilde{\delta}$ fields, which is an entirely new independent object from the original, $\tilde{\Phi} = \tilde{\delta}(\Phi)$. We use the convention that a tilde tensor is equal to the $\tilde{\delta}$ transformation of the original tensor when all its indexes are covariant. This means that $\tilde{S}_{\mu \nu \alpha ...} \equiv \tilde{\delta}\left(S_{\mu \nu \alpha ...}\right)$ and we raise and lower indexes using the metric $g_{\mu \nu}$. Therefore:

\begin{eqnarray}
\tilde{\delta}\left(S^{\mu}_{~ \nu \alpha ...}\right)
&=& \tilde{\delta}(g^{\mu \rho}S_{\rho \nu \alpha ...}) \nonumber \\
&=& \tilde{\delta}(g^{\mu \rho})S_{\rho \nu \alpha ...} + g^{\mu \rho}\tilde{\delta}\left(S_{\rho \nu \alpha ...}\right) \nonumber \\
&=& - \tilde{g}^{\mu \rho}S_{\rho \nu \alpha ...} + \tilde{S}^{\mu}_{~ \nu \alpha ...},
\end{eqnarray}

where we used that $\delta(g^{\mu \nu}) = - \delta(g_{\alpha \beta})g^{\mu \alpha}g^{\nu \beta}$.\\

\subsection*{$\tilde{\delta}$ Transformation:}

With the previous notation in mind, we can define how a tilde component transform. In general, we can represent a transformation of a field $\Phi_i$ like:

\begin{eqnarray}
\bar{\delta} \Phi_i = \Lambda_i^j(\Phi) \epsilon_j,
\end{eqnarray}

where $\epsilon_j$ is the parameter of the transformation. Then $\tilde{\Phi}_i = \tilde{\delta}\Phi_i$ transforms:

\begin{eqnarray}
\label{tilde trans general}
\bar{\delta} \tilde{\Phi}_i = \tilde{\Lambda}_i^j(\Phi) \epsilon_j + \Lambda_i^j(\Phi) \tilde{\epsilon}_j,
\end{eqnarray}

where we used that $\tilde{\delta}\bar{\delta} \Phi_i = \bar{\delta}\tilde{\delta} \Phi_i = \bar{\delta}\tilde{\Phi}_i$ and $\tilde{\epsilon}_j = \tilde{\delta} \epsilon_j$ is the parameter of the new transformation. These extended transformations form a close algebra \cite{AppendixA}.\\

Now, we consider general coordinate transformations or diffeomorphism in its infinitesimal form:

\begin{eqnarray}
\label{xi 0}
x'^{\mu} &=& x^{\mu} - \xi_0^{\mu}(x) \nonumber \\
\bar{\delta} x^{\mu} &=& - \xi_0^{\mu}(x),
\end{eqnarray}

where $\bar{\delta}$ will be the general coordinate transformation from now on. Defining:

\begin{eqnarray}
\label{xi 1}
\xi_1^{\mu}(x) \equiv \tilde{\delta} \xi_0^{\mu}(x)
\end{eqnarray}

and using (\ref{tilde trans general}), we can see a few examples of how some elements transform:\\

\textbf{I)} A scalar $\phi$:

\begin{eqnarray}
\label{scalar}
\bar{\delta} \phi &=& \xi^{\mu}_0 \phi_{, \mu} \\
\label{scalar_tild}
\bar{\delta} \tilde{\phi} &=& \xi^{\mu}_1 \phi,_{\mu} + \xi^{\mu}_0 \tilde{\phi},_{\mu}.
\end{eqnarray}

\textbf{II)} A vector $V_{\mu}$:

\begin{eqnarray}
\label{vector}
\bar{\delta} V_{\mu} &=& \xi_0^{\beta} V_{\mu, \beta} + \xi_{0, \mu}^{\alpha} V_{\alpha} \\
\label{vector_tild}
\bar{\delta} \tilde{V}_{\mu} &=& \xi_1^{\beta} V_{\mu, \beta} + \xi_{1, \mu}^{\alpha} V_{\alpha} + \xi_0^{\beta} \tilde{V}_{\mu, \beta} + \xi_{0, \mu}^{\alpha} \tilde{V}_{\alpha}.
\end{eqnarray}

\textbf{III)} Rank two Covariant Tensor $M_{\mu \nu}$:

\begin{eqnarray}
\label{tensor}
\bar{\delta} M_{\mu \nu} &=& \xi^{\rho}_0 M_{\mu \nu, \rho} + \xi_{0,\nu}^{\beta} M_{\mu \beta} + \xi_{0,\mu}^{\beta} M_{\nu \beta} \\
\label{tensor_tild}
\bar{\delta} \tilde{M}_{\mu \nu}  &=& \xi^{\rho}_1 M_{\mu \nu, \rho} + \xi_{1, \nu}^{\beta} M_{\mu \beta} + \xi_{1, \mu}^{\beta} M_{\nu \beta} + \xi^{\rho}_0 \tilde{M}_{\mu \nu, \rho} + \xi_{0, \nu}^{\beta} \tilde{M}_{\mu \beta} + \xi_{0, \mu}^{\beta} \tilde{M}_{\nu \beta}.
\end{eqnarray}

These new transformations are the basis of $\tilde{\delta}$ Theories. Particulary, in gravitation we have a model with two fields. The first one is just the usual gravitational field $g_{\mu \nu}$ and the second one is $\tilde{g}_{\mu \nu}$. Then, we will have two gauge transformations associated to general coordinate transformation. We will call it extended general coordinate transformation, given by:

\begin{eqnarray}
\label{trans g}
\bar{\delta} g_{\mu \nu} &=& \xi_{0 \mu ; \nu} + \xi_{0 \nu ; \mu} \\
\label{trans gt}
\bar{\delta} \tilde{g}_{\mu \nu} ( x ) &=& \xi_{1 \mu ; \nu} + \xi_{1\nu ; \mu} + \tilde{g}_{\mu \rho} \xi_{0, \nu}^{\rho} + \tilde{g}_{\nu \rho} \xi^{\rho}_{0, \mu} + \tilde{g}_{\mu \nu,\rho} \xi_0^{\rho},
\end{eqnarray}

where we used (\ref{tensor}) and (\ref{tensor_tild}). Now, we can introduce the $\tilde{\delta}$ Theories.\\

\subsection*{Modified Action:}

In the last section, the extended general coordinate transformations were defined. So, we can look for an invariant action. We start by considering a model which is based on a given action $S_0[\phi_I]$ where $\phi_I$ are generic fields, then we add to it a piece which is equal to a $\tilde{\delta}$ variation with respect to the fields and we let $\tilde{\delta} \phi_J = \tilde{\phi}_J$, so that we have:

\begin{eqnarray}
\label{Action}
S [\phi, \tilde{\phi}] = S_0 [\phi] +  \int d^4x \frac{\delta S_0}{\delta \phi_I(x)}[\phi] \tilde{\phi}_I(x),
\end{eqnarray}

the index $I$ can represent any kinds of indices. (\ref{Action}) give us the basic structure to define any modified element for $\tilde{\delta}$ type theories. In fact, this action is invariant under our extended general coordinate transformations developed previously. For this, see \cite{AppendixA}.\\

A first important property of this action is that the classical equations of the original fields are preserved. We can see this when (\ref{Action}) is varied with respect to $\tilde{\phi}_I$:

\begin{eqnarray}
\label{Eq_phi}
\frac{\delta S_0}{\delta \phi_I(x)}[\phi] = 0.
\end{eqnarray}

Obviously, we have new equations when varied with respect to $\phi_I$. These equations determine $\tilde{\phi}_I$ and they can be reduced to:

\begin{eqnarray}
\label{Eq_phi_tilde}
\int d^4x \frac{\delta^2 S_0}{\delta \phi_I(y) \delta \phi_J(x)}[\phi] \tilde{\phi}_J(x) = 0.
\end{eqnarray}

\newpage

\section*{Acknowledgements.}

The work of P. Gonz\'alez has been partially financed by Beca Doctoral Conicyt $N^0$ 21080490, Fondecyt 1110378, Anillo ACT 1102, Anillo ACT 1122 and CONICYT Programa de Postdoctorado FONDECYT $N^o$ 3150398. The work of J. Alfaro is partially supported by Fondecyt 1110378, Fondecyt 1150390, Anillo ACT 1102 and Anillo ACT 11016. J.A. wants to thank F. Prada and R. Wojtak for useful remarks.\\



\begin{thebibliography}{99}
\addcontentsline{toc}{chapter}{Bibliography.}

\bibitem{DM DE 1} D. Hooper and E.A. Baltz, \textit{Strategies for Determining the Nature of Dark Matter}, Annual Review of Nuclear and Particle Science, Vol. 58, Pp. 293-314, (2008), doi:10.1146/annurev.nucl.58.110707.171217.

\bibitem{DM New 1} A. Bosma, Ph.D. thesis, Groningen University (1978) Bib. Code: 1978PhDT.......195B.

\bibitem{DM New 2} A. Bosma and P.C. van der Kruit, \textit{The local mass-to-light ratio in spiral galaxies}, Astronomy and Astrophysics, Vol. 79, Number 3, Pp. 281-286, (1979), Bib. Code: 1979A\&A....79..281B.

\bibitem{DM New 3} V. C. Rubin, W. K. Jr. Ford and N. Thonnard, \textit{Rotational properties of 21 SC galaxies with a large range of luminosities and radii, from NGC 4605 /R = 4kpc/ to UGC 2885 /R = 122 kpc/}, Astrophysical Journal, Part 1, Vol. 238, Pp. 471-487, (1980), doi:10.1086/158003.

\bibitem{DM New 4} P. Salucci, and G. Gentile, \textit{Comment on "Scalar-tensor gravity coupled to a global monopole and flat rotation curves"}, Phys. Rev. D, Vol. 73, Issue 12, 128501, (2006), doi:http://dx.doi.org/10.1103/PhysRevD.73.128501.

\bibitem{DM New 5} P. Salucci, A. Lapi, C. Tonini, G. Gentile, I. A. Yegorova, and U. Klein, \textit{The universal rotation curve of spiral galaxies – II. The Dark Matter distribution out to the virial radius}, MNRAS, Vol. 378, Issue 1, Pp. 41-47 (2007), doi:10.1111/j.1365-2966.2007.11696.x.

\bibitem{DM New 6} M. Persic, and P. Salucci, \textit{The universal galaxy rotation curve}, Astrophysical Journal, Part 1, Vol. 368, Pp. 60-65, (1991), doi:10.1086/169670.

\bibitem{DM New 7} M. Persic, and P. Salucci, \textit{Dark and visible matter in spiral galaxies}, MNRAS, Vol. 234, Issue 1, Pp. 131-154, (1988), doi:10.1093/mnras/234.1.131.

\bibitem{DM New 8} K. M. Ashman, \textit{Dark Matter in galaxies}, Astronomical Society of the Pacific, Vol. 104, Number 682, Pp. 1109-1138, (1992), doi:10.1086/133099.

\bibitem{DM DE 2} A.G. Riess, et al., \textit{Observational Evidence from Supernovae for an Accelerating Universe and a Cosmological Constant}, The Astronomical Journal, Vol. 116, Number 3, Pp. 1009, (1998), doi:10.1086/300499.

\bibitem{DM DE 3} S. Perlmutter, et al., \textit{Measurements of $\Omega$ and $\Lambda$ from 42 High-Redshift Supernovae}, ApJ, Vol. 517, Number 2, Pp. 565, (1999), doi:10.1086/307221.

\bibitem{DM DE 4} R.R. Caldwell and M. Kamionkowski, \textit{The Physics of Cosmic Acceleration}, Annual Review of Nuclear and Particle Science, Vol. 59, Pp. 397-429, (2009), doi:10.1146/annurev-nucl-010709-151330.

\bibitem{DM DE 5} J.A. Frieman, M.S. Turner and D. Huterer, \textit{Dark Energy and the Accelerating Universe}, Annual Review of Astronomy and Astrophysics, Vol. 46, Pp. 385-432, (2008), doi:10.1146/annurev.astro.46.060407.145243.

\bibitem{DM DE 6} M. Milgrom, \textit{A modification of the Newtonian dynamics as a possible alternative to the hidden mass hypothesis}, Astrophysical Journal, Part 1, Vol. 270, Pp. 365-370, (1983), doi:10.1086/161130. Research supported by the U.S.-Israel Binational Science Foundation.

\bibitem{DM DE 7} J. Bekenstein, \textit{Relativistic gravitation theory for the modified Newtonian dynamics paradigm}, Phys. Rev. D, Vol. 70, 083509, (2004), doi:http://dx.doi.org/10.1103/PhysRevD.70.083509.

\bibitem{lambda problem} J. Martin, \textit{Everything you always wanted to know about the cosmological constant problem (but were afraid to ask)}, Comptes Rendus Physique, Vol. 13, Issue 6, Pp. 566-665, (2012), doi:10.1016/j.crhy.2012.04.008.

\bibitem{Deflect} E.B. Formalont and R.A. Sramek, \textit{Measurements of the Solar Gravitational Deflection of Radio Waves in Agreement with General Relativity}, Physical Review Letter, Vol. 36, 1475, (1976), doi:http://dx.doi.org/10.1103/PhysRevLett.36.1475.

\bibitem{delta gravity} J. Alfaro, P. Gonz\'alez and R. \'Avila, \textit{A finite quantum gravity field theory model}, Class. Quantum Grav, Vol. 28, 215020, (2011), doi:10.1088/0264-9381/28/21/215020.

\bibitem{tHooft} G.'t Hooft and M. Veltman, \textit{One-loop divergencies in the theory of gravitation}, Annales de l'institut Henri Poincar\'e. Section A. Physique th\'eorique. Tome 20, num. 1, Pp. $69$-$94$, (1974), http://www.numdam.org/item?id=AIHPA\_1974\_\_20\_1\_69\_0.

\bibitem{Alfaro 2} J. Alfaro, Bv Gauge Theories, (1997), arXiv:hep-th/9702060.

\bibitem{Alfaro 3} J. Alfaro and P. Labra\~na, \textit{Semiclassical gauge theories}, Phys. Rev. D, Vol. 65, 045002, (2002), doi:http://dx.doi.org/10.1103/PhysRevD.65.045002.

\bibitem{DG DE} J. Alfaro, \textit{Delta-gravity and dark energy}, Physics Letters B, Vol. 709, Issues 1-2, Pp. 101-105, (2012), doi:10.1016/j.physletb.2012.01.067.

\bibitem{Paper DE} J. Alfaro and P. Gonz\'alez, \textit{Cosmology in delta-gravity}, Class. Quantum Grav, Vol. 30, 085002, (2013), doi:10.1088/0264-9381/30/8/085002.

\bibitem{Paper 1} J. Alfaro and P. Gonz\'alez, \textit{$\tilde{\delta}$ Gravity, $\tilde{\delta}$ Matter and the accelerated expansion of the Universe}, (2017), arXiv:1704.02888.

\bibitem{Paper 2} J. Alfaro and P. Gonz\'alez, \textit{Non-Relativistic $\tilde{\delta}$ Gravity: A Description of Dark Matter}, (2017), arXiv:1711.00361.

\bibitem{phantom 1} R.R. Caldwell, M. Kamionkowski and N.N. Weinberg, \textit{Phantom Energy: Dark Energy with $\omega < -1$ Causes a Cosmic Doomsday}, Physical Review Letters, Vol. 91, Issue 7, 071301, (2003), doi:http://dx.doi.org/10.1103/PhysRevLett.91.071301.

\bibitem{phantom 2} R.R. Caldwell, \textit{A phantom menace? Cosmological consequences of a dark energy component with super-negative equation of state}, Physics Letters B, Vol. 545, Issues 1-2, Pp. 23–29, (2002), doi:10.1016/S0370-2693(02)02589-3.

\bibitem{phantom 3} S. Nojiri and S. D. Odintsov, \textit{Quantum de Sitter cosmology and phantom matter}, Physics Letters B, Vol. 562, Issues 3-4, Pp. 147-152, (2003), doi:10.1016/S0370-2693(03)00594-X.

\bibitem{phantom 4} J. M. Cline, S. Jeon and G. D. Moore, Physical Review D, \textit{The phantom menaced: Constraints on low-energy effective ghosts}, Vol. 70, 043543, (2004), doi:http://dx.doi.org/10.1103/PhysRevD.70.043543.

\bibitem{phantom 5} G. W. Gibbons, \textit{Phantom Matter and the Cosmological Constant}, DAMTP-2003-19, (2008), arXiv:hep-th/0302199v1.

\bibitem{phantom 6} J. Hao and X. Li, \textit{Attractor solution of phantom field}, Phys. Rev. D 67, 107303, (2017), doi:https://doi.org/10.1103/PhysRevD.67.107303.

\bibitem{massless geo} W. Siegel, \textit{Fields}, Section IIIB, (2005), arXiv:hep-th/9912205.

\bibitem{Landau} L.D. Landau and E.M. Lifshitz, \textit{The Classical Theory of Fields}, Fourth Revised English Edition. Course of Theoretical Physics Volume 2. Institute for Physical Problems, Academy of Sciences of the U.S.S.R. Translated from the Russian by Morton Hamermesh, University of Minnesota. Butterworth-Heinenann. Chapter 10.

\bibitem{Weinberg grav} S. Weinberg, \textit{Gravitation and Cosmology: Principles and Applications of the General Theory of Relativity}, Massachusetts Institute of Technology (1972). See Chapter 8.

\bibitem{dark matter prop} N. P. Pitjev and E. V. Pitjeva, \textit{Constraints on Dark Matter in the Solar System}, Astron. Lett. (2013) 39: 141. doi:10.1134/S1063773713020060.

\bibitem{AppendixA} See Appendix A from: J. Alfaro, \textit{Delta-gravity, Dark Energy and the accelerated expansion of the Universe}, J. Phys.: Conf. Ser, Vol. 384, 012027, (2012), doi:10.1088/1742-6596/384/1/012027.
\end{thebibliography}
\end{document}